# Observed Quantization of Anyonic Heat Flow

Mitali Banerjee[+], Moty Heiblum[+], Amir Rosenblatt[+], Yuval Oreg[+], Dima E. Feldman[*], Ady Stern[+], and Vladimir Umansky[+]

[+] *Broun Center for Sub-Micron Research, Dep. of Condensed Matter Physics,*
*Weizmann Institute of Science, Rehovot, Israel 76100*

[*] *Department of Physics, Brown University, Providence, Rhode Island 02912, USA*

## ABSTRACT

The quantum of heat conductance of ballistic one-dimensional (1D) channels, being $g_Q=\kappa_0 T$ with $\kappa_0=\pi^2 k_B^2/3h$ ($T$ - temperature, $k_B$ - Boltzmann's constant, $h$ - Planck's constant), is an important fundamental constant. While the quantization of the electrical conductance of 1D ballistic conductors has long been experimentally established, a demonstration of the quantization of thermal conductance proved to be much harder. It has already been accomplished for weakly interacting systems of phonons, photons and electronic Fermi-liquids. Theoretically, however, the quantization must also hold in strongly interacting systems, such as the Fractional Quantum Hall effect (FQHE), where electrons fractionalize into anyons and chargeless quasiparticles such as neutral Majorana fermions. Since the bulk in the FQHE is incompressible, it is not expected to contribute significantly to the heat conductance, which is determined by chiral 1D edge modes. The thermal conductance reflects topological properties of the FQH electronic systems to which the electrical conductance gives no access. Here, we present results of extensive measurements of the heat conductance in 'particle-like' (Laughlin's) and 'hole-like' fractional states. We verify the universal value of the quantum of thermal conductance of the charged fractional modes as well as for chargeless neutral modes. We also prove the validity of the theoretical predictions for the more complex (and less studied) 'hole-like' states. Heat transport measurements open a door to ample information, not easily accessible by conductance measurements.

**Introduction**

The Fractional Quantum Hall (FQH) state, observed first in 1982, still provides a plethora of challenges [1]. The universal quantized Hall conductance, with current flowing in chiral one-dimensional (1D) edge modes, is directly related to the bulk Landau level filling $v$, $G_H=dI/dV=vG_0$, with $G_0=e^2/h$ being the quantum conductance ($I$ - current, $V$ - voltage, $e$ - electron charge, $h$ – Planck's constant). The nature and number of the edge modes is not dictated by topological considerations, and may take different values for different FQH states at the same Landau level filling [2,3]. Consequently, the electrical conductance reflects the number and the conductance of charged chiral modes, but is blind to the total number of the modes, their respective chirality, and their character.

The thermal conductance, $G_Q=dJ_Q/dT$, with $J_Q$ being the heat current, is intimately connected with the **net** number of the edge modes (*downstream* minus *upstream*) – independent of their quasiparticle charge [3,4]. This is a direct consequence of information-theoretic arguments, put forward by John Pendry in 1983, that a single 'quantum channel' could conduct heat only up to a universal maximum value determined by a quantum of thermal conductance $g_Q=\kappa_0 T$, with $\kappa_0=\pi^2 k_B^2/3h$ ($k_B$ – Boltzmann's constant) [5]. This quantization has already been experimentally verified in weakly-interacting systems, including systems of phonons [6,7], electrons [8,9,10], and photons [11].

Here, we extend the study of thermal transport to a strongly - interacting system provided by the FQHE [1]. Our purpose is to test the notion that the quantum of heat conductance is universal and independent of the quasiparticles charge; which is a fraction of the electron charge or even zero. Specifically, we studied non-interacting electrons (in the Integer QHE), a 'particle-like' (Laughlin's) state, and, the more intriguing and complex, 'hole-like' states - in which the FQHE liquid is formed by holes in a filled Landau level. After verifying the quantization of thermal transport in the IQHE, we studied first the $\nu=1/3$ Laughlin's state, with $G_H=(1/3)G_0$ and quasiparticle charge $e^*=e/3$, which is expected to carry one unit of the quantum conductance ($G_Q=\mathbf{1}\cdot\kappa_0 T$, without factor 1/3). Alternatively, 'hole-states' with filling $1/2<\nu<1$, which were already shown to support downstream charge and upstream modes [12], are thus expected to have a 'topological' thermal conductance determined by the **net** chirality of all their edge modes. For example, the $\nu=2/3$ state is described by downstream charge mode (or more – see later) with the total conductance $G_H=(2/3)G_0$ and upstream chargeless modes [2,3,12,13,14,15]. These topological modes might be augmented by pairs of added counter-propagating charged modes. Since the number of *downstream* modes must be equal to that of the *upstream* ones, the thermal transport is diffusive and the **net** thermal conductance goes to zero at large system sizes [2,3]. Other interesting states are $\nu=3/5$ and $\nu=4/7$, with the **net** heat current expected to propagate, remarkably, in the **opposite** direction to the charge current [3,16].

**Methodology**

*The basic idea*

Our experimental setup adopts the core idea put forward by Jezouin *et al.* [6], who measured the heat conductance in the IQHE. Here, however, we exploit a more flexible implementation shown in Fig. 1 (see below). A DC input power is provided by a voltage source $V_S$ that drives current $I_S=G_H V_S$. The current carries the power $P_S=0.5 I_S V_S$ (half of the incoming power is dissipated at the *upstream* side of the Source contact). QPC1 determines the impinging current at the floating contact reservoir, $I_{in}=t_1 I_S$, where $t_1=\nu_{QPC1}/\nu$ is the transmission on the conductance plateau at $G_{QPC1}=\nu_{QPC1}G_0$. The outgoing current from the reservoir splits into $N$ arms (here, $N_{max}=4$), carrying the power $P_{out}=P_{in}/N$, with the dissipated power in the reservoir $\Delta P=P_{in}-P_{out}=0.5 I_{in} V_S (1-N^{-1})$. The system reaches equilibrium, in which the dissipated power leaves the reservoir via phonons (to the bulk underneath and to the arms) and via the downstream and upstream edge modes, $\Delta P=\Delta P_e+\Delta P_{ph}$. The reservoir reaches equilibrium at a temperature $T_m>T_0$, where $T_0$ is the electron temperature in all other contacts. Consequently, the chiral *ballistic* 1D edge modes are expected to carry heat $\Delta P_e = 0.5 \cdot N \cdot \kappa_o \cdot (T_m^2 - T_0^2)$, while the heat flux via phonons is expected to obey $\Delta P_{ph} = \beta (T_m^5 - T_0^5)$ [17]. We find that the latter becomes negligible at $T_m < 35$mK in comparison to the electronic contribution. The temperature $T_m$ is determined by the thermal noise measured (by a cooled pre-amplifier) in one or two of the *N* arms.

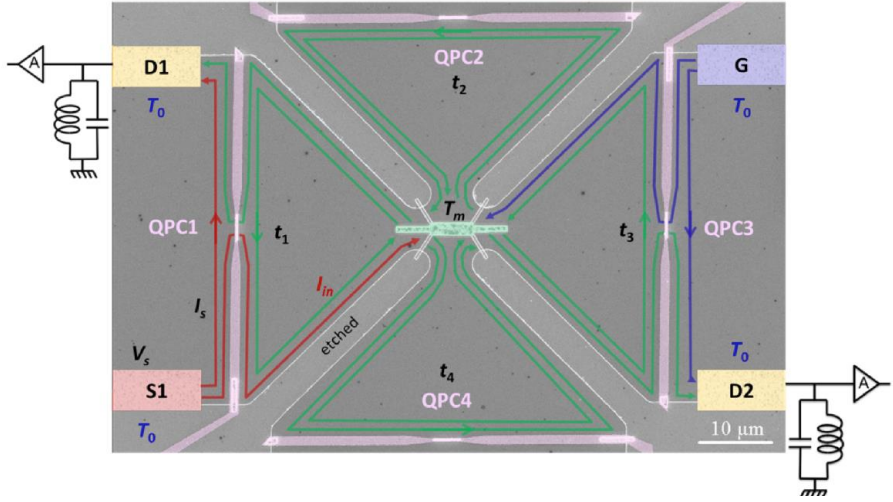

**Figure 1. Configuration of the actual device**. An SEM micrograph of the four-arm device, with a small floating ohmic contact at its center (green; the depleting groves underneath are not visible), and a quantum point contact (QPC) in each arm (an air-bridge shorts the two sides of the spit-gate). The Source (**S1**), Drains (**D1, D2**) and ground (**G**) contacts are drawn not to scale. In this example, $v=2$ and QPC2 and QPC4 are fully pinched (thus $N=2$) while QPC1 and QPC3 transmit only the outmost edge mode and fully reflect the inner mode. The Source current ($I_S$, red) impinges on QPC1, which transmits $I_{in}$ (here $I_{in}=I_S/2$) that is absorbed in the floating contact. Edge modes (green) at a potential $V_m$ and at temperature $T_m$, leave the floating contact into the four arms (in arms *N*2 and *N*4 they are fully reflected). Cold edge modes, at temperature $T_0$ (blue) arrive from the grounded contacts. An *LC* circuit at each drain filters the signal ($f_0=740$kHz and band-width $\Delta f=10$-$30$kHz depending on $v$), which is amplified by the voltage pre-amplifier (cooled to 4.2K), followed by a room-temperature amplifier (total gain ~1000). The amplified signal is measured by a spectrum analyzer with similar $f_0$ and $\Delta f$.

In a general case of multiple 1D edge modes in each of the *N* arms, the expressions for the dissipated power and the noise are a bit more cumbersome. We express the dissipated power in the floating contact as:

$$\Delta P = \frac{1}{2} \frac{I_S^2}{G_0} \frac{v_{QPC1}}{v^2} \left[ 1 - v_{QPC1} \bigg/ \sum_{i=1}^{i=4} v_{QPCi} \right].$$ In turn the temperature $T_m$ is related to the thermal noise via

$S_{th} = 2G^* k_B (T_m - T_0)$, with $\dfrac{1}{G^*} = \dfrac{1}{G_{amp}} + \dfrac{1}{\sum\limits_{i=1, i \neq amp}^{i=4} G_i}$, where $G_i$ is the conductance of the *i*'th arm [18,19,20].

*The actual realization*

An SEM micrograph of the 'heart' of the device is shown in Figs. 1. The floating small contact (light green) was utilized in two different sizes, 'small' - ~8μm$^2$ in a 'high' density 2DEG, and 'large' - ~18μm$^2$ in a 'low' density 2DEG. Four arms were formed by chemical etching; each with a QPC that can be partly or fully be pinched. The Source contact S1 is located in arm #1 with the 'cold pre-amplifier' (cooled to 4.2K) located in the opposite arm. The amplifier (calibrated by shot-noise measurements) measured the thermal voltage fluctuations on the edge mode $S_υ$ after they were filtered by an *LC* circuit ($f_0$~740kHz and BW=$\Delta f$=10-30kHz depending of $v$). The desired current fluctuations were calculated via $S_{th}=S_υ G_H^2$.

The 'high' density 2DEG, with $n_e$=1.1x10$^{11}$cm$^{-2}$ and 4.2K mobility μ=4x10$^6$cm$^2$/V-s, was used in the Integer regime; while the 'low' density 2DEG, with $n_e$=0.88x10$^{11}$cm$^{-2}$ and 4.2K mobility μ=4.6x10$^6$cm$^2$/V-s, was used in the fractional regime. Since the density tends to drop near edge of the contacts, a larger size contact was made in the lower density material in order to minimize reflection [21]. Etched groves under the floating contact (not visible in the micrograph), made sure that the impinging current enters the ohmic contact before splitting to the different arms.

*Points of consideration*

Deducing the thermal conductance with a reasonable accuracy necessitates a careful determination of the system's parameters, such as the amplification of the amplifiers-chain, reflection from the floating small contact, and the temperatures $T_0$ and $T_m$. Doing that, we assume the following: (*i*) Excess shot noise is not produced by the floating small contact, because: 1. The reflection coefficient of the contact is small enough (we find <2%); 2. Full equilibration takes place in the contact; 3. Charging of the contact is negligible as its *RC* time is very short; (*ii*) The phonon contribution to the heat dissipation into the bulk and to the arms is independent of the bulk filling factor and the number of the participating arms. (*iii*) The arms are long (~150μm to the Ground contact), thus allowing equilibration of counter-propagating modes (however, the critical length is not known); and (*iv*) Bulk energy modes, though weak, are possible [15].

**Measurements**

*The integer regime*

We start with the integer regime, with $v$=1 & 2 [10], where the 1D chiral edge modes are known to flow downstream with no counter-propagating modes [12]. We elaborate here on $v$=2 [see $v$=1 in Supplementary Section]. Two edge modes leave the Source, while 1 or 2 modes impinge at the floating contact depending on QPC1. A few tests were performed first, leading to: (*i*) The Source does not produce excess noise; (*ii*) The reflection coefficient from the floating contact is negligible; (*iii*) The current splits equally between all the open arms. The excess thermal noise $S_{th}$, was then measured as a function of the Source current $I_S$ in a few different fillings in the QPCs, $v_{QPCi}$ (Fig. 2a). Using the above expressions, $T_m$ was calculated as a function of the dissipated power $\Delta P$ (for $T_0$=30mK, Fig. 2b). Since in these measurements the temperature was relatively high, the phonon contribution was subtracted as demonstrated in Fig. 2c; namely, $\delta P=\Delta P(n_i,T_m)-\Delta P(n_j,T_m)$, and with its normalized form $\lambda=\delta P/(\kappa_o/2)$ plotted in six different configurations of $\Delta N=n_i-n_j$ ($n_i$ is the total number of outgoing modes). The average thermal conductance for the 1D mode is found to be $g_Q$=**(0.98±0.032)**$\kappa_o T$; in good agreement with the expected quantization. We also find the 'phonon coefficient' $\beta$ for that floating contact, $\beta \sim$(3-5)nW-K$^{-5}$. Similar measurements were performed in $\nu$=1 with a smaller number of configurations. Here, the average thermal conductance per 1D mode was $g_Q$=**(0.9±0.09)**$\kappa_o T$ (Fig. S2).

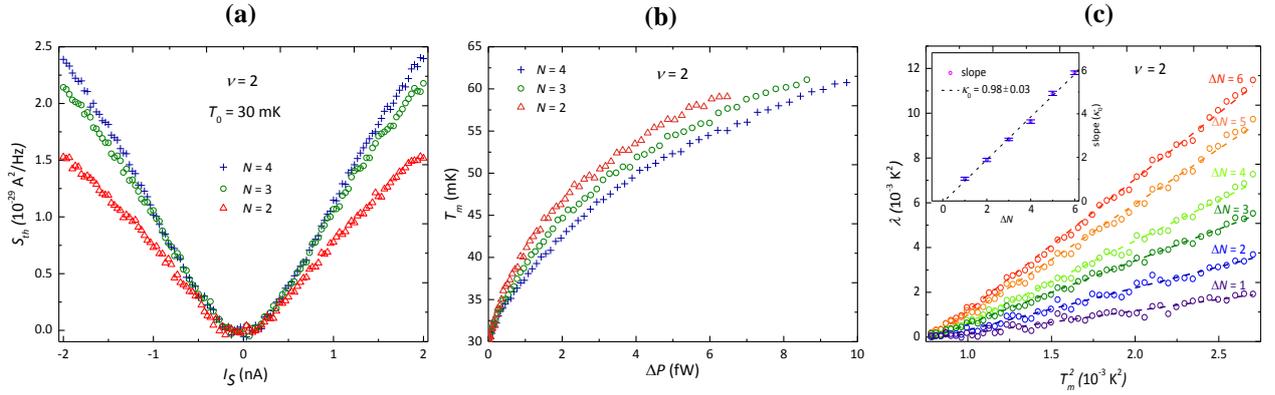

**Figure 2. Measurements in the integer QHE at filling $\nu$=2.** **(a)** Excess thermal noise $S_{th}$ plotted as a function of the Source current $I_S$ in three different configurations, where only the outmost mode is transmitted through the QPCs. **(b)** The calculated temperature of the floating contact $T_m$ plotted as a function of the dissipated power in the latter three configurations. **(c)** Subtracting dissipated powers at different $n$'s ($n$ the total number of modes) at $T_m$ allows eliminating the phonons contribution. Plotted are fits of normalized $\delta P=\Delta P(n_i,T_m)-\Delta P(n_j,T_m)$ by $\kappa_o/2$, which we name $\lambda$, as function of $T_m^2$ for six different combinations of $n_i$ and $n_j$. **Inset.** The slopes of each of the six combinations (red circles) and a line-fit (dotted black line), provides an average thermal conductance of $g_Q$=**(0.98±0.03)**$\kappa_o T$ for each 1D mode.

### *The fractional regime - $\nu$=1/3 particle-state*

The $\nu$=1/3 state is the most heavily researched state. Being the first filled Landau level of Composite Fermions $\nu_{CF}$=1, with $\nu=\nu_{CF}/(2\nu_{CF}+1)$ [22], it is known to harbor a single chiral downstream edge mode (when edge potential is soft, edge-reconstruction may add pairs of counter-propagating neutral modes [15]). Hence, its heat conductance is expected theoretically to be $g_Q$=**1**·$\kappa_o T$ – like for $\nu$=1. We studied this state in the lower density sample with the larger size floating contact (the measured reflection was ~2%, leading to an insignificant shot noise in each of the outgoing modes), and at $T_0$~10mK. The power dissipation was kept small enough to avoid a large temperature increase, allowing neglecting the phonons contribution. Source noise, reduced by a factor of $N^2$, as subtracted from the measured thermal noise. As both the conductance and temperature range were small, the thermal noise $S_{th}$=$2G^*k_B(T_m-T_0)$ was also small, leading to a somewhat larger uncertainty in the data.

We plotted in Fig. 3a the dissipated power as function of $T_m^2$ for two configurations, $N=4$ and $N=2$, and the theoretically expected dependence. The absence of a $T_m^5$ term verifies that the phonon contribution is negligible. In Fig. 3b we plot the normalized subtracted powers, $\lambda$, for $\Delta N=(N=4)-(N=2)$ as function of $T_m^2$, in order to reassure the elimination of the phonon contribution. The average thermal conductance is $g_Q=(\mathbf{1.0\pm0.045})\kappa_o T$; again, in excellent agreement with the expectations.

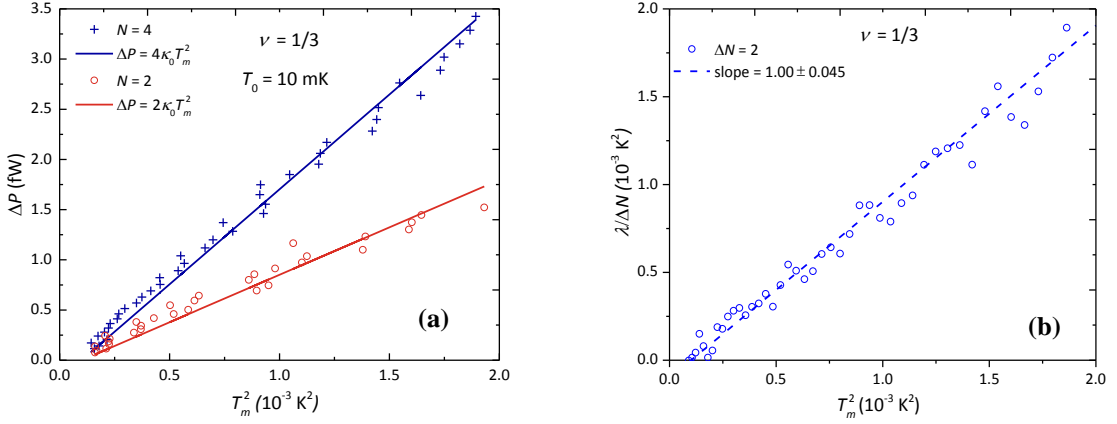

**Figure 3. Measurements in the fractional QHE at filling $v=1/3$. (a)** Total power carried out by the fractional edge modes from the floating contact for two different configurations ($N=4$ blue and $N=2$ red). The theoretical expectation (with no phonons contribution at this temperature range), $\Delta P_e=0.5\cdot N\kappa_o T_m^2$, is plotted as function of $T_m^2$. The RMS of the data points deviates by 4% at $N=4$ and 2% at $N=2$. **(b)** Plotting $\lambda$, the normalized $\delta P=\Delta P(N4,T_m)-\Delta P(N2,T_m)$ by $\kappa_o/2$, as function of $T_m^2$ (thus subtracting the phonon contribution, if any - purple dots), the fit line gives $g_Q=(\mathbf{1.00\pm0.045})\kappa_o T$.

*The fractional regime - hole-states*

The so-called 'hole-states', with fractional fillings ½<$v$<1, belong to the same composite fermion hierarchy as the 'particle-states' (however, the residual magnetic field left after flux attachment is opposite in direction to the original field, and hence edge modes propagate in both chiralities). These states are related to the actual electron filling via $v=v_{CF}/(2v_{CF}-1)$. Here, we study the states at $v=2/3$, $v=3/5$, and $v=4/7$, with quantized conductance plateaus, $G_H=vG_0$. Due to counter-propagating modes [3,12], their thermal conductance is not obvious; it is theoretically predicted to be: $g_Q=\mathbf{0}$ at $v=2/3$, and quantized as $g_Q=\mathbf{-1}\kappa_o T$ at $v=3/5$ and as $g_Q=\mathbf{-2}\kappa_o T$ at $v=4/7$. Astonishingly, in the latter two, heat is expected to propagate **opposite** (negative sign) to the direction of the charge current [3,16]. Note, however, that the actual sign of the **net** chirality could not be determined in our device configuration.

In our experiment, the downstream modes leave the floating contact with its electrochemical potential and temperature $T_m$ on one edge of each arm, while the upstream neutral modes emanate from the contact at the opposite edge of the arm – also heated by $T_m$. Hence, both downstream and upstream modes take heat away from the contact. Charge and neutral modes arrive back from the grounded contacts (blue lines in Fig. 1), intermix and equilibrate with the downstream modes, and thus bring energy back to the floating contact.

The initial ballistic propagation of downstream modes turns diffusive at long enough propagation length, and heat dissipation via the edge modes is restricted.

The 'hole-like' states appeared to be 'experimentally friendly'. In particular, the dip in the density near the periphery of the contacts does not lead to excess noise. We attribute that to an increase in the CF filling as the density lowers (filling approaches $v=1/2$), allowing for a smooth traversal of the lower - CF filling edge modes [22].

We performed similar measurements in the three 'hole-like' states. First we measured the total power carried out from the floating contact by the fractional edge modes for $N=4$, and then the normalized subtracted powers, $\lambda$, for $\Delta N=(N=4)-(N=2)$, as function of $T_m^2$; in order to assure the elimination of phonon contribution. It appears that the propagation length is sufficiently long to allow equilibration between the counter-propagating modes, which is necessary in order to obtain the quantized thermal conductance.

The $v_{CF}=2$, $v=2/3$ state. As we expect the state to support an equal number of downstream and upstream modes [3], its thermal conductance should vanish ($g_Q\sim 0$) for sufficiently long propagation length (neglecting phonons). While our understanding today is that the state supports two downstream $G_0/3$ modes and two upstream neutral modes [23], the **net** chirality is still zero and thus the thermal conductance is expected to be very small. We measured the partitioned downstream charge in a partly pinched QPC (say, with the conductance of the QPC, $G_0/3=G_H/2$) to be $e^*=2e/3$ (as in Ref. 24) (Fig. S3).

Total out-flowing electronic power for $N=4$ is plotted as function of $T_m^2$ (Fig. 4a, $v=2/3$ black). The data deviates from the expected small value. In Fig. 4b we plot the normalized value of the subtracted powers at $N=4$ and $N=2$, $\lambda$, as function of $T_m^2$. In both measurements, we find a non-zero thermal conductance, $g_Q\approx(\mathbf{0.33\pm 0.024})\kappa_o T$. This finite value of heat conductance decreased to $g_Q\approx \mathbf{0.25}\kappa_o T$ when the temperature increased $T_0=30$mK. This result might be related to a significant heat flow through the bulk [15], which is missing in our model. Another possibility could be related to a relatively long equilibration length, allowing thus that part of the energy outflow be ballistic without equilibration. In both cases, we expect sensitivity to temperature.

The $v_{CF}=3$, $v=3/5$ state. This state is expected to have a **net** upstream chiral mode in the limit of a long propagation length; with the heat carried by that mode – as all the other modes flow diffusely and may carry only a small amount of heat away from the floating contact. Though the total number of modes is not known, we found by gradually pinching a QPC that the state supports at least two downstream charge modes; an outmost one with conductance $G_0/3=5G_H/9$ and an inner one with conductance $2G_0/5=2G_H/3$. These modes are evidently accompanied by upstream neutral modes [12]. The partitioned quasiparticle charge was measured to be $e^*=3e/5$ at 10mK (Fig. S3).

As above, we plot in Fig. 4a the power dissipation for $N=4$ followed by $\lambda$, the normalized $\delta P$, in Fig. 4b. We obtained the average thermal conductance $g_Q=(\mathbf{1.04\pm 0.03})\kappa_o \Delta T$, which agrees with our expectations. Increasing the temperature to $T_0=30$mK did not alter the result. It appears that all the edge modes, except **one**, become fully diffusive, and thus do not carry any heat away from the floating contact even at the lowest temperature.

<u>The $v_{CF}=4$, $v=4/7$ state.</u> Here, we found three chiral edge modes near the edge: from the outmost channel with conductance $G_0/3$, the middle with $2G_0/5$, and innermost with $3G_0/7$ (Fig. S4). The partitioned quasiparticle charge was found to be $e^*=4e/7$ at 10mK (Fig. S3). Repeated measurement in this fractional state led to an average of the thermal conductance $g_Q=(\mathbf{2.045\pm0.05})\kappa_o\Delta T$ (Fig. 4a & 4b); reconfirming the theoretical prediction.

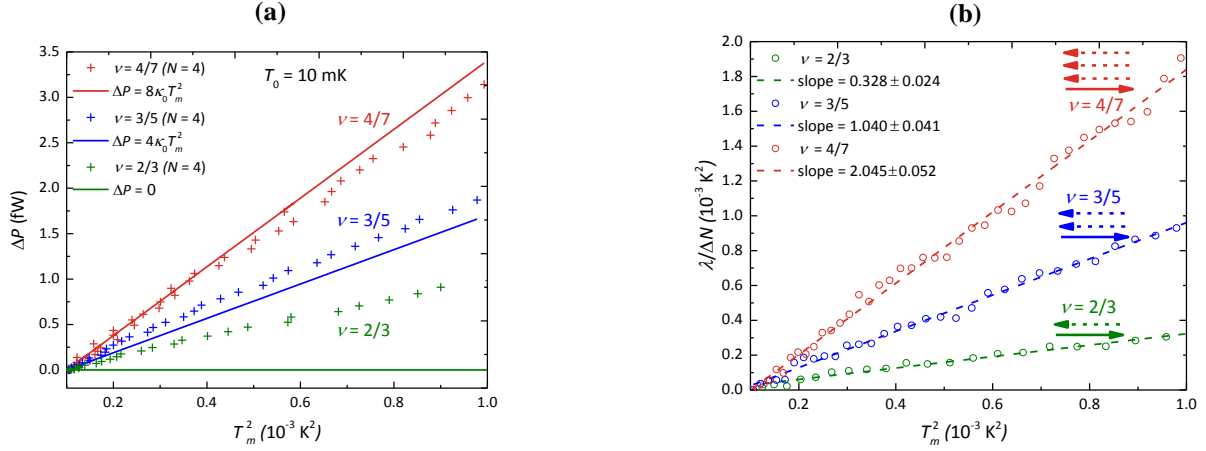

**Figure 4. Measurements in the fractional QHE at hole-like states.** **(a)** Total power carried out from the floating contact by the fractional edge modes for $N=4$ in the three hole-like states as function of $T_m^2$ ($v=2/3$ black, $v=3/5$ blue, $v=4/7$ red). The expected theoretical result, $\Delta P_e=0.5\cdot 4\cdot n\kappa_o T_m^2$ with $n\kappa_0$ being the expected thermal conductance; namely, $n=0$ for $v=2/3$, $n=-1$ for $v=3/5$, and $n=-2$ for $v=4/7$, are plotted in solid lines. The RMS of the data points deviates from the expected values by 12% for $v=3/5$ and by 8.5% for $v=4/7$. Deviation for in the case of v=2/3 is large (see text). **(b)** Plotting $\lambda$, the normalized subtracted dissipated powers $\delta P=\Delta P(N4,T_m)-\Delta P(N2,T_m)$ by $\kappa_o/2$, as function of $T_m^2$ (thus subtracting the phonon contribution, if any), the fit lines agrees quite well with the predictions for the thermal conductance, except for the $v=2/3$ state. Arrows show the minimal edge models for each filling factor: Solid arrows depict *downstream* charged modes and dashed arrows depict *upstream* neutral modes.

## Discussion

The extraordinarily precise quantization of the electrical Hall conductance is a striking example of a topological phenomenon in physics. Yet, the conductance is just one signature of the topological order and different orders may exhibit the same conductance. The measurement of topological properties that distinguish between such orders is a major challenge. The experiment we report on in this paper successfully addresses this challenge by measuring the thermal Hall conductance.

Beyond the obvious difficulties associated with the measurement of thermal Hall conductance, our measurements indicate that the topological protection of the quantization of the thermal Hall conductance is weaker than that of the electrical Hall conductance. While the localization of the bulk states suppresses charge transfer, it does not suppress energy transfer at the same efficiency. An example to that is the Wigner crystal, which is electrically insulating in the DC limit but may transfer heat through its phonons.

We describe experimental studies of six different quantum Hall states; two integer states (at filling $v=1$ & 2) – that set the background for this work, an 'electron-like' $v=1/3$ fractional state, and three 'hole-like' states, $v=2/3$, $v=3/5$ and $v=3/7$. We verified the value of the quantum of the heat conductance in these systems. While for the integer states and the $v=1/3$ state, the heat is carried by downstream modes, the hole-like states are more complex as heat is carried by counter-propagating modes - some charged and some neutral. While the presence of neutral modes was already verified before, the energy current they carry was not known. Our results are consistent with the fundamental theory that predicts that these chargeless modes carry the same heat as the charged ones (irrespective of their velocities and the interaction with the other chiral modes or with the bulk).

An exciting feature of this type of experiments is that they can now be extended to other poorly understood quantum Hall states. For example, there are important questions about the second Landau level, which may host a family of non-abelian states (such as $v=5/2$ and $v=12/5$). Some of the proposed topological orders may serve as a platform for universal (topological based) quantum computations. Thus far, these states remained enigmatic with no definite proof of their nature. The thermal conductance would provide a *smoking gun* evidence of their topological order.

**Author contributions**

M.B and A.R. designed the experiment, and fabricated the devices. A.R. participated in initial measurements. M.B. and M.H. preformed the measurements, did the analysis and guided the experimental work. Y.O., D.F, & A.S. worked on the theoretical aspects. V.U. grew the actual 2DEG heterostructures. All contributed to the write up of the manuscript.

**Acknowledgments**

M.B. acknowledges the help and advice pf R. Sabo and I. Gurman. M.H. acknowledges the European Research Council under the European Community's Seventh Framework Program (FP7/2007-2013)/ERC Grant agreement No. 339070, the partial support of the Minerva foundation, grant no. 711752, and together with V.U., the German Israeli Foundation (GIF), grant no. I-1241-303.10/2014. A.S and Y.O. acknowledge the European Research Council under the European Community's Seventh Framework Program (ERC) Grant agreement No. 340210, and the Israeli Science Foundation, ISF agreement No. 13335/16. D.E.F. acknowledges support by the NSF under Grant No. DMR-1205715.

## Supplementary Section

Here we provide more detailed information on experimental details and add more examples of raw results.

*Fabrication*

- **2DEG** Two structures were grown.

'High density' - $n_e$=1.1x10$^{11}$cm$^{-2}$, 4.2K mobility $\mu$=4x10$^6$cm$^2$/V-s,

2DEG depth below surface 113nm, spacer 80nm, quantum well width 30nm.

'Low density' - $n_e$=0.88x10$^{11}$cm$^{-2}$, 4.2K mobility $\mu$=4.6x10$^6$cm$^2$/V-s,

2DEG depth below surface 128nm, spacer 95nm, quantum well width 30nm.

- **Ohmic contacts** Evaporation sequence, from surface of the GaAs and up:

Ni (5nm) Ge (100nm) Au (200nm) Ni (75nm) Au (150nm). Alloyed at 450C for 50sec.

- **QPCs** Evaporation sequence from GaAs and up: Ti (5nm) Au (20nm).

Split gate gap: (1) 'high density' - 700nm; 'low density' - 850nm.

- **Air Bridges** Evaporation sequence, from surface of the GaAs and up: Ti (20nm) Au (480nm). Two-layer resist process.

*Characterization*

- **QPCs** The QPCs were cooled while biased at +0.3V gate voltage. Yet, at composite fractions, such as $v$=2/5 or $v$=4/3, the inner edge mode was partly reflected. Moreover, at 'zero bias cooling', the QPCs reflected a large fraction of the total current.

- **Source noise** Though the transmission of the contact was high (with small series resistance), some ohmic contacts produced noise at high magnetic field (fillings $v$<1/2), which could be related to a lower density near their periphery. The Source noise magnitude was ~1.5x10$^{-29}$A$^2$/Hz at the highest Source current used. This was divided for subtraction purposes by 4, 9, and 16, for $N$=2, 3, and 4, respectively.

- **Small floating contact** Reflection was measured by comparing the full reflected current from a pinched QPC to the pre-amplifier, and the reflected current for $N$=2 (or larger). The reflection was less than 2% in all the measured fractional states. Assuming this reflection led to shot noise, its magnitude would have been much smaller than the measured thermal noise. Measuring the expected thermal conductance with the reports accuracy, confirms that the noise is small.

- **Gain calibration** Knowing the gain is crucial for the determination of the actual temperature $T_m$. The gain of the amplification chain and the 'cold temperature' $T_0$ were measured by measuring shot noise (in a side QPC on the same device) in fractional states where the partitioned charge is well known (such as $v$=2, $v$=2/3, and more). Examples of actual shot noise measurements are provided below

- **Branching of current** Equal branching of impinging current into all the $N$ arms is a prerequisite for our measurement scheme to be successful. In order to verify this, we have measured the transmission of each QPC, while the other arms QPCs are either open or closed. For example, if we source and measure in the same arm, and if all the other QPCs are open, then the measured signal in the Source arm will start at relative transmission of 0.25 when the QPC is fully open, and unity when the QPC is fully closed. When one arm will close, the transmission in the Source arm will drop to 0.33. Another configuration of measurement is to source in one arm and measure in another arm.

Below are two examples of measurements done in $v=1/3$ and $v=3/5$.

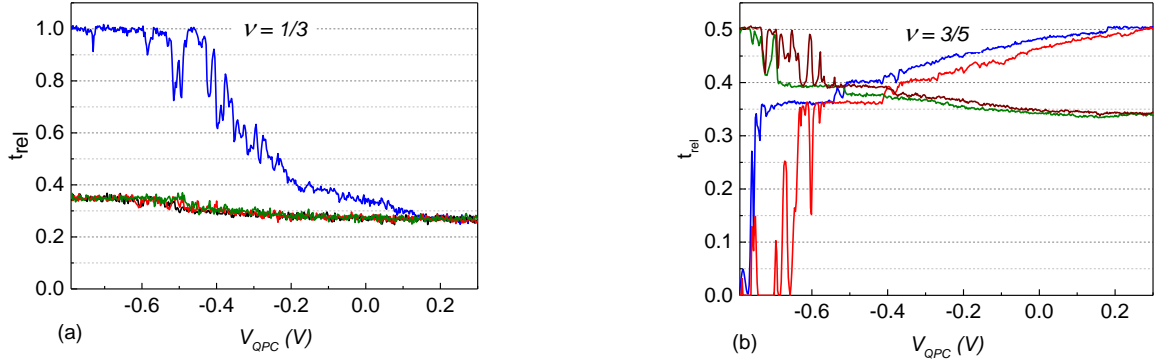

**Fig. S1.** Relative transmission of all the arms' QPCs. **(a)** $v=1/3$. Current sourced from S1 and measured by the amplifier at D1. The blue curve is a function of pinching Source QPC, when all other arms' QPCs are open. The other colored lines are when each of the arms QPCs pinches separately, while Source QPC is open. **(b)** $v=3/5$. Current is sourced from S1 and measured by the amplifier at D2, as the different QPCs pinch. The red and blue lines are the transmissions QPC1 and QPC3 while both QPC2 and QPC4 were completely pinched, so the relative transmission starts from 0.5 and goes down to 0. The other two lines are transmission of QPC2 while all QPCs except QPC4 were open and vice versa. Here the relative transmission starts from 0.33 and reaches 0.5 when QPC2 or QPC4 was completely pinched.

*Added data*

**Data at $v=1$**

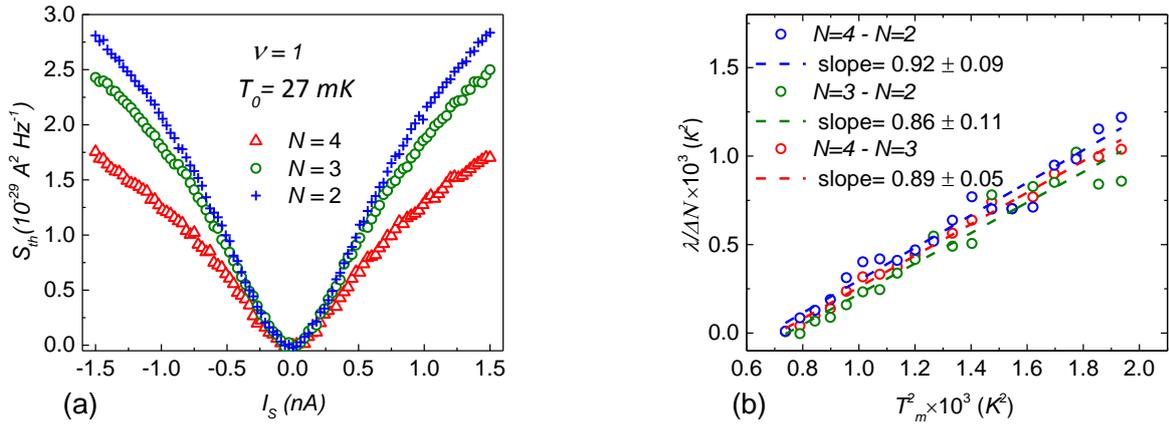

**Fig. S2.** **(a)** Thermal noise as a function of $I_s$ at different arms (different $N$) configurations in $v=1$ at $T_0=27$mK. **(b)** Plotting the normalized subtracted power dissipations, noted $\lambda$ in text, as function of $T_m^2$ (thus subtracting the phonon contribution up to $T_m=40$mK). The fit line gives $g_Q=(0.9\pm0.09)\kappa_o T$.

## Data for 'hole-like' states

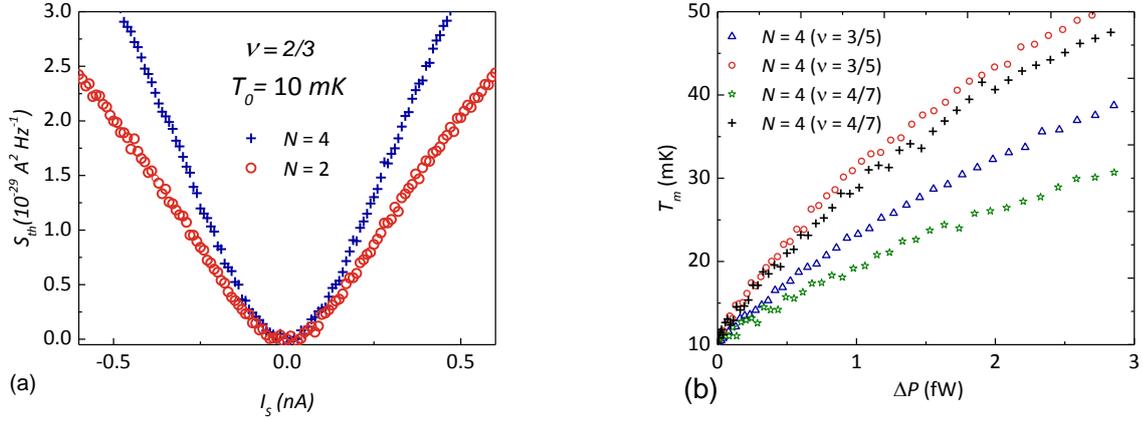

**Fig. S3. (a)** Thermal noise in $v=2/3$ is plotted for two different arm configurations for $T_0$=10mK. **(b)** Plotting $T_m$ as function of dissipated power for both $v=3/5$ and $v=4/7$, in both cases $T_0$=10mK.

## Conductance of QPC and shot noise in 'hole-like' states

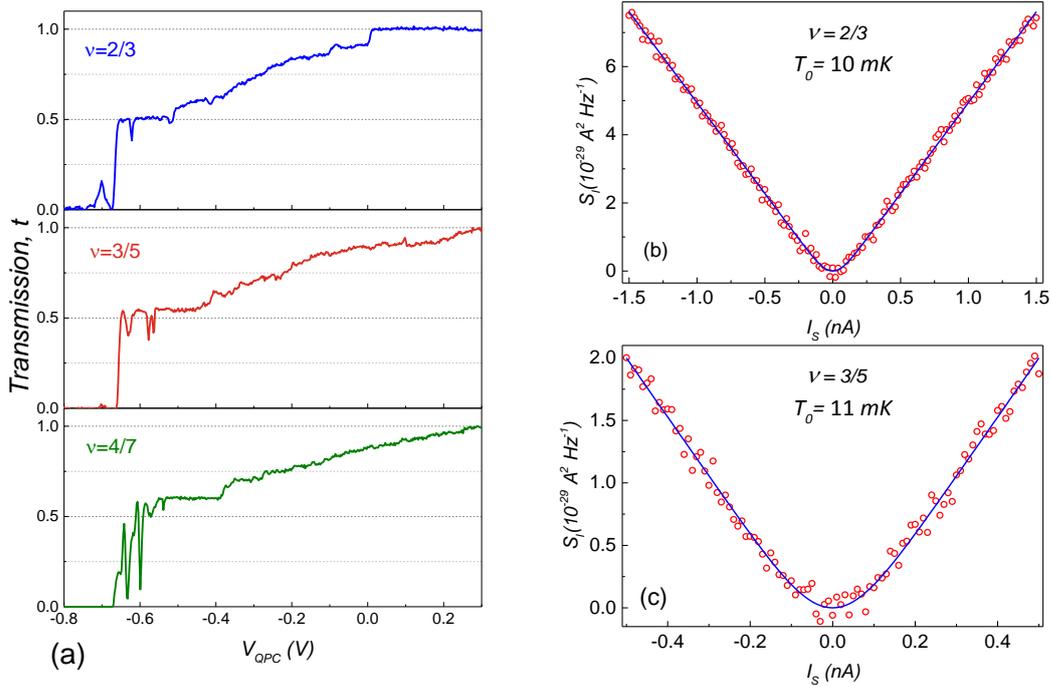

**Fig. S4. (a)** Transmission of QPC at $v=2/3$, 3/5 and 4/7. Plateaus are for the different chiral charge modes at the edge. **(b)** Shot noise measured $v=2/3$ in order to calibrate the gain at a transmission of the QPC $t$=0.5. The solid line shows the fit with the equation $2e^*It(1-t)[\coth(e^*V/2k_BT)-2k_BT/e^*V]$, with $e^*=2e/3$ and $T$=10mK. **(c)** Shot noise measured at $v=3/5$, with $t$=0.55, with the solid line shows the fit to $e^*=3e/5$ and $T$=11mK.